\documentclass[aps,prl,epsf,showpacs,amsmath,amssymb,graphics,twocolumn,10pt]{revtex4}
\usepackage{graphicx}
\usepackage{dcolumn}
\usepackage{bm}

\newcommand{\be}{\begin{equation}}
\newcommand{\ee}{\end{equation}}
\newcommand{\ba}{\begin{eqnarray}}
\newcommand{\ea}{\end{eqnarray}}
\newcommand{\ban}{\begin{eqnarray*}}
\newcommand{\ean}{\end{eqnarray*}}

\newcommand{\eq}[1]{(\ref{#1})}

\begin{document}

\title{Naked Singularities as Particle Accelerators}

\author{Mandar Patil \footnote{mandarp@tifr.res.in}
and  Pankaj S. Joshi \footnote{psj@tifr.res.in}}

\affiliation{Tata Institute of Fundamental Research\\
Homi Bhabha Road, Mumbai 400005, India}


\begin{abstract} We investigate here the particle
acceleration by naked singularities to arbitrarily high center
of mass energies. Recently it has been suggested that black holes
could be used as particle accelerators to probe the Planck scale
physics. We show that the naked singularities serve the same
purpose and probably would do better than their black hole
counterparts. We focus on the scenario of a self-similar
gravitational collapse starting from a regular initial data, leading
to formation of a globally naked singularity. It is seen that
when  particles moving along timelike geodesics interact and
collide near the Cauchy horizon, the energy of collision
in the center of mass frame will be arbitrarily high,
thus offering a window to Planck scale physics.

\end{abstract}

\maketitle

Recently, an interesting observation was made by Banados,
Silk and West
\cite{BSW},
that black holes can accelerate infalling colliding particles to arbitrarily
high energy in the center of mass frame around the horizon of an
extremal Kerr black hole, provided certain restrictive conditions
were imposed on the angular momenta of the particles.
While this mechanism was criticized for being a very fine
tuned one
\cite{Berti},
\cite{Jacobson},
 a number of further works have recently investigated
the process for different black holes, e.g. in the context of
an extremal charged spinning black hole
\cite{Wei},
non-extremal rotating black holes
\cite{Grib1, Grib2},
non-rotating charged black-holes
\cite{Zasla2},
and so on. It is also claimed that this may be a generic property of
rotating black holes in a model independent way
\cite{Zasla1}.

We show here that the divergence of center of
mass energy of colliding particles is a phenomenon not only
associated with black holes, but also with naked singularities
which are outcome of a continued gravitational collapse
of a massive star. We consider here a self-similar spherically
symmetric spacetime. There have been several numerical and
analytical investigations of self-similar gravitational collapse
for different matter fields satisfying reasonable energy
conditions, such as dust
\cite{dust},
ideal fluids with non-vanishing pressures
\cite{Ori},
massless scalar fields
\cite{scalar1, scalar2},
and such others, leading to the formation of naked singularities
in gravitational collapse from a regular initial data. Naked singularity
formation in collapse models  has
been investigated in detail in recent years (see e.g.
\cite{Joshi}
and references therein), and a wide variety of physically
relevant situations, self-similar or otherwise, are included.

Whether or not naked singularities would occur
in the real world we live in, is an unanswered question till this
date. However, considering the very many results of
gravitational collapse scenarios that lead to a naked singularity
final fate for collapse, we may assume that naked singularities
could occur in various physical circumstances such as the final fate
of a massive star, when it collapses at the end of its life cycle on
exhausting its internal nuclear fuel. It would be then of much
interest to investigate the consequences of their formation,
taking into account the possible quantum gravity effects
these may cause
\cite{Goswami},
and possible connection with the very highly energetic
astrophysical phenomena such as the gamma ray bursts
and those related to the active galactic nuclei.

We show here that the naked singularities forming in
gravitational collapse could provide us a
window into the new Planck scale physics, even far away
from the actual singularity. This is because, unlike the case of
a black hole, the particles could go very close to a naked singularity
and then emerge with very high velocities near the Cauchy horizon
that the naked singularity created.  For definiteness, we focus
here on self-similar models but the results may hold good
in more general class of collapses.

A self-similar spacetime is characterized by the presence of
a vector field $\xi$, known as the homothetic killing vector, which satisfies
\begin{equation}
\emph{L}_{\xi}g_{\mu\nu}=2g_{\mu\nu}
\label{homothetic}
\end{equation}
where $\emph{L}_{\xi}$ stands for a Lie derivative along
the vector field $\xi$, and $g_{\mu\nu}$ is the spacetime metric.
A spherically symmetric self-similar spacetime geometry
in the $\left(t,r,\theta,\phi\right)$ coordinates
can be written as,
\begin{equation}
ds^2=-e^{2\nu\left(X\right)}dt^2+e^{2\psi\left(X\right)}dr^2+
r^2S\left(X\right)^2d\Omega^2 \label{metric}
\end{equation}
where $\xi=t\partial_{t}+r\partial_{r}$ is the homothetic vector field
and  $X= t/r$ is the self-similarity variable, $d\Omega^2$ being the
metric on a two-sphere. It is easily verified that the metric
\eq{metric} satisfies \eq{homothetic}.
For a given matter field, one can write down and solve the
Einstein field equations, which reduce to ordinary differential equations
in this case, to obtain the metric functions
$\nu\left(X\right)$,$\psi\left(X\right)$,and $S\left(X\right)$.
As mentioned, the exact solution to the Einstein field
equations for dust as a matter field
\cite{dust},
and numerical solution with an ideal fluid with linear
equation of state
\cite{Ori}
have been obtained and are shown to admit the
naked singularity as collapse final state.
We keep our analysis here general and model
independent, without explicitly referring to any of the
particular solutions mentioned above, though clearly
our conclusions apply to each of these cases.

The naked singularity as final outcome of collapse is
characterized by existence of families of outgoing null and timelike
geodesics, which terminate in the past at the singularity and
in future they reach a faraway observer in the spacetime.
The first null geodesic that comes out from the singularity
is the Cauchy horizon in the spacetime.
Before considering the particle collissions near the
Cauchy horizon, we first note certain general features of
a self-similar spacetime admitting naked singularity in
collapse, developing from a regular initial data.

The Ricci scalar for \eq{metric} is of the form,
\begin{equation}
\Re = \frac{f\left(X\right)}{r^2}
\label{ricci}
\end{equation}
where $f\left(X\right)$ is a function of self-similarity
variable alone, which can be written explicitly in terms
of the metric functions $\nu\left(X\right)$,$\psi\left(X\right)$, and
$S\left(X\right)$. If $f(X_{0})$ is a finite, non-zero number,
approaching $\left(t=0,r=0\right)$ along the curve $t=X_{0}r$ makes
the Ricci scalar diverge as $r\rightarrow 0$. Hence $t=0,r=0$
is the spacetime singularity. The full singularity curve is
given by $\frac{t}{r}=X=X_{s}$, where $X_{s}$ is a smallest
positive number for which $f\left(X_{s}\right)\rightarrow \infty$
and the Ricci scalar \eq{ricci} diverges. It is known that
various other curvature scalars also diverge at the singularity
$t=0,r=0$, which is a strong curvature sinularity in
the spacetime.

We are interested in the situation where gravitational
collapse starts from a regular data specified on an initial
spacelike hypersurface. Therefore $f\left(X\right)$ is finite for
$X<0$, i.e. $t<0$. The regular initial data implies the regularity
of center $r=0$ for $t<0$, i.e. as $X\rightarrow -\infty$.
Hence $f\left(X\right)\sim \frac{1}{X^2}$ as $X\rightarrow -\infty$,
so that $\Re\sim \frac{1}{t^2}$, and the center $r=0$ is
regular for $t<0$. The structure of such a collapse
spacetime is depicted in Fig 1.

\begin{figure}
\begin{center}
\includegraphics[width=0.4\textwidth]{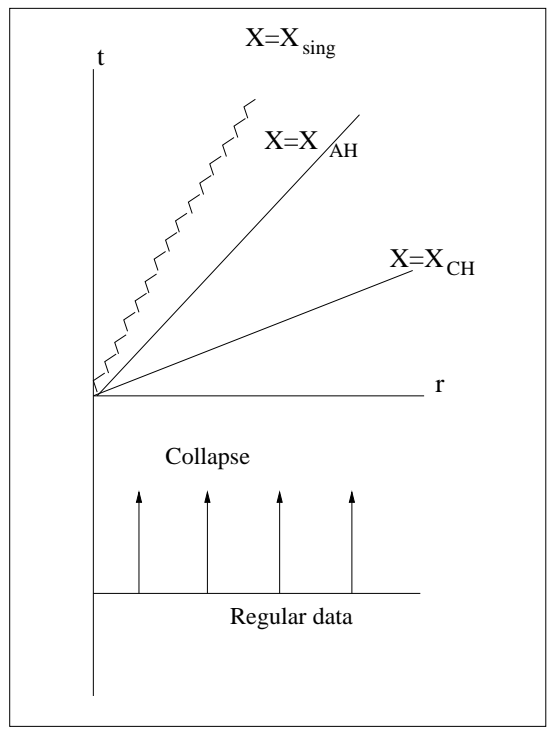}
\caption{\label{fg1}
The general structure of a self-similar spherically symmetric
spacetime admitting naked singularity. The singularity curve, apparent
horizon, and the Cauchy horizon are given by $X=X_{sing},X_{ah},X_{ch}$
respectively.}
\end{center}
\end{figure}

The apparent horizon in the spacetime is given
by $g^{\mu\nu}\partial_{\mu}R\partial_{\nu}R=0$, where
$R=rS\left(X\right)$, and has the equation,
\begin{equation}
e^{2\psi}S_{X}^2=e^{2\nu}\left(S-XS_{X}\right)^2
\label{ah1}
\end{equation}
The solution $X=X_{ah}$ to \eq{ah1} gives the apparent
horizon curve ${t}/{r}=X_{ah}$. If $X_{s}<X_{ah}$ then it would
be possible to have outgoing null and timelike curves from
the singularity curve to infinity, which would be globally visible.
When the central singularity $\left(t=0,r=0\right)$ is visible,
the equation of outgoing radial null geodesics $ds^2=0$
from the same is given by $\left(\frac{dt}{dr}=e^{\psi-\nu}\right)$.
Since close to the $\left(t=0,r=0\right)$, $\frac{dt}{dr}\approx \frac{t}{r}=X$,
we get,
\begin{equation}
X^2e^{2\nu}=e^{2\psi}
\label{cauchy}
\end{equation}
close to the origin.  However, it is easily verified
in this case that this also represents a null geodesic away from the
origin. There would be outgoing radial null geodesics to
infinity from $\left(t=0,r=0\right)$ if there exists a positive solution
$X=X_{c}$ to \eq{cauchy}, such that $X_{c}<X_{ah}$, and the singularity
would then be naked. We then have a situation where a self-similar
spherically symmetric gravitational collapse from a regular initial data
on an initial hypersurface gives rise to a naked singularity,
and there are outgoing null geodesics to infinity from
$\left(t=0,r=0\right)$. The equation \eq{cauchy} also describes
the Cauchy horizon, which is the first outgoing null geodesic
from the naked singularity.

Consider now a test particle of mass $m$, following a
timelike radial geodesic on this spacetime. Let $\lambda$ be
an affine parameter along the geodesic and $U^{\mu}$ be
the velocity vector. Then using the geodesic equation
$U^{\mu};_{\nu}U^{\nu}=0$ and normalization condition
$U^{\mu}U_{\mu}=-1$, we get,
\begin{equation}
\frac{d}{d\lambda} \left(U^{\mu}\xi_{\mu}\right)=-1
\label{geodesic1}
\end{equation}
Using \eq{metric} we obtain
\begin{equation}
\left(U^{\mu}\xi_{\mu}\right)=-e^{2\nu}tU^{t}+e^{2\psi}rU^{r}=C-\lambda
\label{geodesic2}
\end{equation}
where $C$ is an integration constant. The normalization
condition for velocity is
\begin{equation}
-e^{2\nu}\left(U^{t}\right)^2+e^{2\psi}\left(U^{r}\right)^2=-1
\label{normalization2}
\end{equation}
The nonvanishing components of velocity are,
\begin{equation}
U^{t}=\frac{\left(X\pm e^{2\psi}Q\right) \left( C-\lambda\right)}
{r \left( e^{2\psi}-e^{2\nu}X^2\right)};
U^{r}=\frac{\left(1 \pm Xe^{2\nu} Q\right) \left( C-\lambda\right)}
{r \left( e^{2\psi}-e^{2\nu}X^2\right)}
\label{geodesic3}
\end{equation}
where
\[
Q=\sqrt{e^{-2\psi-2\nu}+\frac{r^2 e^{-2\psi-2\nu}\left(e^{2\psi}-e^{2\nu}X^2\right)}
{\left(C-\lambda\right)^2}}
\]
Here the positive and negative signs correspond to
outgoing and ingoing geodesics respectively
\cite{Joshi}\cite{Joshi2}.
It can be verified easily that the above satisfy
\eq{geodesic2},\eq{normalization2}.

Consider now two particles of identical mass $m$ (for simplicity),
which collide very close to the Cauchy horizon, the first being an outgoing
one, coming from a close vicinity of the naked singularity along a radial
timelike geodesic from the singularity, and the second being an infalling
one that crosses the Cauchy horizon to fall inwards.
From \eq{cauchy},
since $e^{2\psi}-X^2e^{2\nu}\approx 0$, $Q\approx e^{-\psi-\nu}$,
so the components of velocities (outgoing $U^{\mu}_{+}$ and
ingoing $U^{\mu}_{-}$) of particles can be written as,
\begin{equation}
U^{t}_{+}=\frac{B_1 e^{-\nu}}{r \left(e^{\psi}-Xe^{\nu}\right)} ;
U^{r}_{+}=\frac{B_1 e^{-\psi}}{r \left(e^{\psi}-Xe^{\nu}\right)}
\label{outgoing}
\end{equation}
\begin{equation}
U^{t}_{-}=\frac{-B_2 e^{-\nu}}{r \left(e^{\psi}+Xe^{\nu}\right)} ;
U^{r}_{-}=\frac{B_2 e^{-\psi}}{r \left(e^{\psi}+Xe^{\nu}\right)}
\label{ingoing}
\end{equation}
where $B_i = C_i-\lambda_{i}$. We focus here only on the outgoing
radial timelike geodesics in the region of spacetime beyond the Cauchy
horizon where $e^{2\psi}-X^2e^{2\nu}<0$.
Since $U^{t}>0$, we must have $B_1 <0 , B_2 <0$.

The energy of collision in the center of mass frame
is given by \cite{BSW},
\begin{equation}
E_{cm}=\sqrt{2}m \sqrt{1-2g_{\mu \nu} U_{1}^{\mu}U_{2}^{\nu}} \label{ECM}
\end{equation}
The center of mass energy computed from ingoing
and outgoing particles of mass $m$, colliding close to the Cauchy horizon,
from \eq{ingoing},\eq{outgoing},\eq{ECM} is then given by,
\begin{equation}
E_{cm}\approx \frac{2\sqrt{2}m}{r}  \sqrt{\frac{B_1 B_2}
{e^{2\nu} X^2-e^{2 \psi}}} \label{ECM2}
\end{equation}
This would be arbitrarily large, depending on how close
is the point of collision to the Cauchy horizon. A point to be noted
here is that in general the collision event could be far away from
the singularity.

The advantage of the scenario considered here is,
we already have the outgoing timelike geodesics existing
from an arbitrary vicinity of the singularity, by the very structure of
the spacetime geometry.  These then meet the infalling particles
arbitrarily close to the Cauchy horizon. The key difference
from its black hole counter-part is, we consider here the
collision between an outgoing and an ingoing particle, rather
than two ingoing ones as in the black hole case.

Such outgoing particles from a close neighborhood of
singularity, which travel close to the Cauchy horizon, could arise
in various ways. One can consider the region of spacetime
before the formation of singularity, and the ingoing geodesics
starting from a faraway region, after passing through the regular
center would emerge as outgoing geodesics. The outgoing particles
close to the Cauchy horizon would be the ingoing geodesics, which
just missed the singularity $\left(t=0,r=0\right)$, and emerge as
outgoing particles. Such ingoing particles in collapse may miss the
singularity if they had a small angular momentum or due to the
small perturbations in geometry. The region around singularity
would be fuzzy, and dominated by quantum gravity effects.
When non-perturbative semi-classical modifications to the classical
evolution dynamics are taken into account from quantum gravity,
the fuzzy region around what would have been classically a
naked singularity, is shown to be "super-repulsive"
for arbitrary matter field configurations in the late regime
\cite{Singh},\cite{Goswami}.
Hence the particles emerging out of this region may get
boosted up in energy significantly and would travel close to
the Cauchy horizon for a long time. These particles, accelerated
by the singularity, would then meet the ingoing particles close
to the Cauchy horizon, and the collision would occur at arbitrarily
large center of mass energies.


It would be interesting to examine whether this result would
generalize to many of the collapse scenarios where a naked singularity
develops even when self-similarity is relaxed. The
divergence of colliding particles near Cauchy horizon could
then be a generic phenomenon associated with naked singularities.
In that case one could probe the Planck scale physics in a
region away from the actual singularity, but near the Cauchy horizon.
For instance, one may construct a model of an inhomogeneous
dust collapse of a finite matter cloud (Fig.2), the exterior being a Schwarzschild
metric. For a wide class of initial conditions of the initial
density and velocity profiles, the collapse would give rise to a
naked singularity, and eventually a black hole. Before the apparent
horizon engulfs the surface of the star, turning it into a black hole,
the Planck scale physics would be visible at the surface,
as the Cauchy horizon hits it. This might trigger unknown channels
of reactions between elementary particles and could have impact
on our understanding of phenomena such as the active galactic
nuclei, provided they can be modeled this way. Since dark matter
could be modeled by dust, the situation described above may
be applicable to the gravitational collapse of halos of dark matter,
and might have astrophysical implications.

\begin{figure}
\begin{center}
\includegraphics[width=0.4\textwidth]{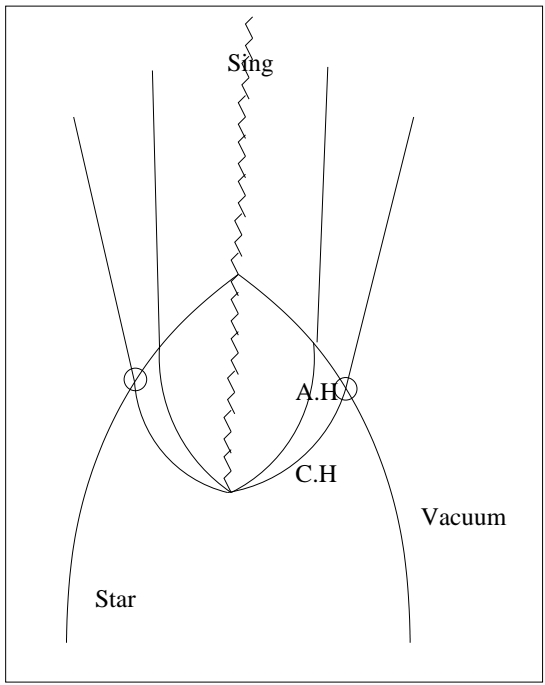}
\caption{\label{fg2}
The matter cloud undergoes a collapse to form a
naked singularity. A.H. is apparent horizon, C.H. is Cauchy horizon,
Sing is singularity. The circle depicts region on the surface
of the star where Planck physics would be visible.}
\end{center}
\end{figure}

It is interesting to note here that the Cauchy horizon
for black holes, which
also coincides with their inner horizon, exhibits an instability
termed as mass inflation
\cite{Poisson}.
The particles colliding near the Cauchy horizon of naked
singularity and exhibiting divergence in the center of mass
energy frame may indicate another form of instability for
the Cauchy horizon. We have treated here the colliding particles
as test particles, ignoring gravitational radiation and
the backreaction on the spacetime.

There are key differences between the particle acceleration
by black holes and naked singularities. The divergence of center of
mass energy in case of a black hole is near its event horizon.
The new particles created in such a collision would either enter
the horizon and would never be detected, or if they escape the black hole
they could be infinitely redshifted for an asymptotic observer
to see them, and for whom the process requires an infinite amount
of time
\cite{Jacobson}.
Hence it is very unlikely that black holes as super-colliders
would be useful to probe Planck scale physics. However, since the
collisions with divergent center of mass energies occur near the
Cauchy horizon in the case of naked singularities, it would be possible to
detect the particles, as the redshift and time required for this process
would be finite. Hence, as a probe of Planck scale physics
naked singularities could do a better job compared to
their black hole counterparts.

\end{document}